\documentclass[11pt,a4paper]{article}
\usepackage[left=2.5cm,right=2.5cm]{geometry}
\usepackage{amsmath,amssymb,amsthm}
\usepackage{amsfonts}
\usepackage{amssymb}
\usepackage{graphicx, rotating}
\usepackage{epstopdf}
\usepackage{epsfig}
\usepackage{latexsym}
\usepackage{graphicx}
\usepackage{color}
\usepackage{amsmath,amssymb}
\usepackage{slashed}
\usepackage{hyperref}
\usepackage{booktabs}
\usepackage{multirow}
\usepackage{hyperref}
\usepackage{bbm}
\usepackage{lscape}
\usepackage{amscd}
\usepackage{feynmp}
\usepackage{placeins}

\makeatletter
\def\endfmffile{%
\fmfcmd{\p@rcent\space the end.^^J%
end.^^J%
endinput;}%
\if@fmfio
\immediate\closeout\@outfmf
\fi
\ifnum\pdfshellescape=\@ne
\immediate\write18{mpost \thefmffile}%
\fi}
\makeatother

\newcommand{\tev}{\mbox{ TeV}}

\newcommand{\mt}{m_{\text{top}}}
\newcommand{\mh}{m_{\text{h}}}
\newcommand{\mkk}{m_{\text{t,KK}}}
\allowdisplaybreaks

\begin{document}
\unitlength=1mm  %%feynmp Diagramgröße festlegen

\begin{titlepage}
\begin{flushright}
\end{flushright}
\vspace{.3in}

\begin{center}
\vspace{1cm}

{\Large \bf
Applying EFT to Higgs Pair Production in Universal Extra Dimensions
}

\vspace{1.2cm}
{\large Lisa Edelh\"auser, Alexander Knochel and Thomas Steeger}
\vspace{.8cm}

{\it {Institut f\"ur Theoretische Teilchenphysik und Kosmologie, RWTH Aachen,
Aachen, Germany}}\\

\begin{abstract}
\medskip
\noindent
We investigate single Higgs and Higgs pair production at the LHC in models of
Universal Extra Dimensions. After calculating the relevant cross
sections, we use the UED model as a testing ground for the Effective Field Theory
approach to physics beyond the Standard Model.  We show how the UED
contributions to Higgs production can be matched to a dimension-6 operator. We
then discuss the range of validity of this approach, in particular for Higgs
pair production, and determine the sensitivity to the number of KK modes in the
loop.  
\end{abstract}
\vspace{.4cm}

\end{center}
\vspace{.8cm}

\end{titlepage}

\section{Introduction}
After the discovery of a Higgs boson at the LHC
\cite{Chatrchyan:2012ufa,Aad:2012tfa}, the precise determination of its
properties is of paramount importance.  Many of the couplings to gauge bosons
and fermions are already being measured in single Higgs production.  However, in
order to discriminate between the weakly interacting Standard Model Higgs and
alternative EWSB scenarios, it is crucial to measure the Higgs
self-interaction~\cite{Baur:2003gp,Baur:2003gpa} which so far has only been
determined indirectly from the Higgs mass in a model-dependent fashion. An
important process allowing a direct measurement of these couplings at the LHC is
Higgs pair production. 

An important difference between on-shell single and double Higgs production via
gluon fusion is the invariant mass flowing through the quark loop at LO, which
is fixed at $p^2=m_h^2$ in the first case\footnote{See however the possibility
to resolve the momentum dependence of the effective $ggh$-Vertex in gluon fusion
via the radiation of additional jets \cite{Buschmann:2014twa} and off-shell
Higgs measurements \cite{Azatov:2014jga}.}, but is only constrained by the
collider energy in the other.  Therefore, double Higgs production can yield
additional sensitivity to new physics contributing to gluon fusion which can
affect the self coupling measurement.  In this work we consider a Standard Model
extension in which the Higgs self coupling remains unchanged at tree level,
while the loop in double Higgs production potentially resolves additional heavy
degrees of freedom. Concretely, we use a minimal Universal Extra Dimensions
scenario \cite{Appelquist:2000nn}, where for the purpose of this work we are
mainly interested in the heavy quark spectrum and its couplings to the lightest
Higgs. 

If one is interested in a model-independent description of heavy new physics in
Higgs production, the effective field theory (EFT) framework using dimension-6
operators
\cite{Buchmuller:1985jz,Elias-Miro:2013mua,Corbett:2013pja,Pomarol:2013zra}
proves to be a very efficient tool.  However, since the EFT expansion becomes
invalid for invariant masses near the cutoff scales~\cite{Biekoetter:2014jwa},
comparisons with concrete UV completions are important in order to understand
the precise scale and nature of this breakdown. We therefore match the new
physics contributions to gluon fusion in our Universal Extra Dimensions (UED)
scenario to the corresponding effective operator(s) and compare the results to
the full 1-loop calculation.  Higgs pair production has been calculated
previously in such models, and we find a discrepancy between both our full
1-loop and EFT results and the published result in \cite{deSandes:2007my}. 
\section{LO Higgs Production in a UED model}
Gluon fusion via (top) quark loops constitutes the dominant single and double
Higgs pair production mode in the SM, and this remains true in the UED extension
we consider here. We take into account the top quark and its Kaluza-Klein (KK)
excitations in the loop.  The concrete scenario we use is a simple version of
mUED in which we neglect the mass splittings from loop effects and related
boundary terms as higher order effects.  Starting point is a 5D version of the
SM Lagrangian compactified on a circle of radius $R$ in which each chiral SM
fermion is extended to a full Dirac spinor. A $\mathbb Z_2$ orbifold symmetry
reduces the extra dimension to an interval of length $y\in [0,\pi R]$ with
orbifold fixed points at the boundaries $y=0,\pi R$. Its main purpose is to
project out lefthanded singlets, righthanded doublets and the $V_5$ components
of gauge bosons from the massless spectrum in order to recover the chiral SM in
the low energy limit.  The doubling of the fermionic field content in the 5D
theory yields two Dirac type partners at each massive KK level for each SM Dirac
fermion. For the top we denote them by $t_1^{n}$ and $t_2^{n}$ respectively.
They are degenerate in mass at tree level.  Details about this construction can
be found in \cite{Appelquist:2000nn,Papavassiliou:2001be}.

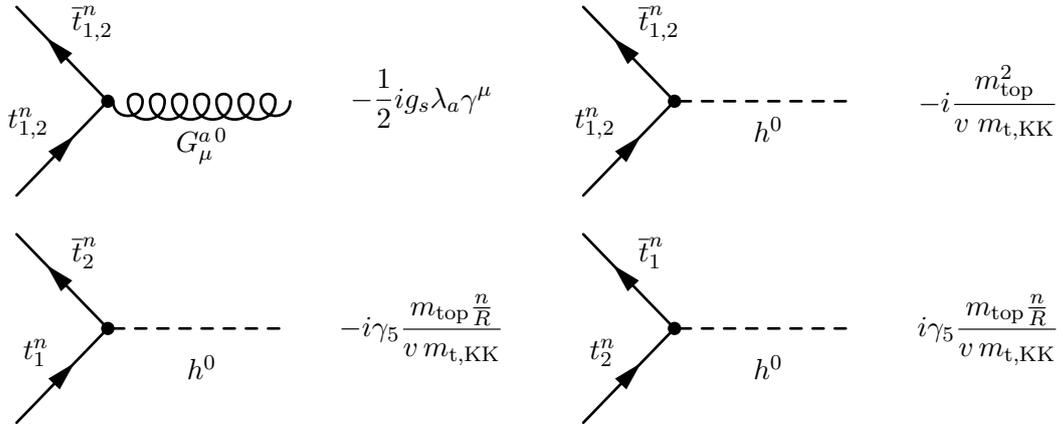
\begin{figure}
\begin{center}
  \begin{minipage}[c]{0.25\textwidth}
 \begin{fmffile}{gghed1}
  \begin{fmfgraph*}(40,25)
   \fmfleft{i1,i2}
   \fmfright{o1}
   \fmf{gluon,label.dist=10,label=$G_\mu^{a \,, 0}$}{v1,o1}
   \fmf{fermion,label=$t_{1,,2}^n$}{i1,v1}
   \fmf{fermion,label=$\overline{t}_{1,,2}^n$}{v1,i2}
   \fmfdot{v1}
  \end{fmfgraph*}
 \end{fmffile}
 \end{minipage}
 \begin{minipage}[c]{0.2\textwidth}
  \begin{align*}
\raisebox{1.1cm}{$\displaystyle 
- \frac{1}{2} i g_{s} \lambda_a \gamma^\mu$}
\end{align*}
 \end{minipage} 
  \begin{minipage}[c]{0.25\textwidth}
 \begin{fmffile}{gghed3}
  \begin{fmfgraph*}(40,25)
   \fmfleft{i1,i2}
   \fmfright{o1}
   \fmf{dashes,label=$h^0$}{v1,o1}
   \fmf{fermion,label=$t_{1,,2}^n$}{i1,v1}
   \fmf{fermion,label=$\overline{t}_{1,,2}^n$}{v1,i2}
   \fmfdot{v1}
  \end{fmfgraph*}
 \end{fmffile}
 \end{minipage}
 \begin{minipage}[c]{0.2\textwidth}
  \begin{align*}
\raisebox{1.1cm}{$\displaystyle 
- i \frac{\mt^2}{v \; m_{\text{t,KK}}}$}
\end{align*}
 \end{minipage}\\ 
\begin{minipage}[c]{0.25\textwidth}
 \begin{fmffile}{gghhed1}
  \begin{fmfgraph*}(40,25)
   \fmfleft{i1,i2}
   \fmfright{o1}
   \fmf{dashes,label.dist=10,label=$h^0$}{v1,o1}
   \fmf{fermion,label=$t_{1}^n$}{i1,v1}
   \fmf{fermion,label=$\overline{t}_{2}^n$}{v1,i2}
   \fmfdot{v1}
  \end{fmfgraph*}
 \end{fmffile}
 \end{minipage}
 \begin{minipage}[c]{0.2\textwidth}
  \begin{align*}
\raisebox{1.1cm}{$\displaystyle 
- i \gamma_5 \frac{\mt \frac{n}{R}}{v \, \mkk}$}
\end{align*}
 \end{minipage}
 \begin{minipage}[c]{0.25\textwidth}
 \begin{fmffile}{gghhed2}
  \begin{fmfgraph*}(40,25)
   \fmfleft{i1,i2}
   \fmfright{o1}
   \fmf{dashes,label.dist=10,label=$h^0$}{v1,o1}
   \fmf{fermion,label=$t_{2}^n$}{i1,v1}
   \fmf{fermion,label=$\overline{t}_{1}^n$}{v1,i2}
   \fmfdot{v1}
  \end{fmfgraph*}
 \end{fmffile}
 \end{minipage}
 \begin{minipage}[c]{0.2\textwidth}
  \begin{align*}
\raisebox{1.1cm}{$\displaystyle 
 i \gamma_5 \frac{\mt \frac{n}{R}}{v \, \mkk}$}
\end{align*}
 \end{minipage}
\end{center}
 \caption{The relevant Feynman rules for Higgs production via gluon fusion in
our UED scenario. Here, $\lambda_a$ denote the usual Gell-Mann matrices, $v$
and $g_s$ are the 4D values of the vacuum expectation value and the strong
coupling respectively. Furthermore, $n\in \mathbb N_0$ denotes the KK mode
number. For $n=0$, which corresponds to the SM particles which obtain all mass
via the Higgs mechanism, only $t=t_1^0$ exists while $t_2^0$ is projected out
by the orbifold.\label{feynmanrules}} \end{figure}

These two degenerate top partners per level $t_1^n$ and $t_2^n$ receive a mass
\begin{equation}\mkk = \sqrt{\mt^2 + n^2/R^2}\,\end{equation} which includes an
SM-like contribution $m_{top}$ from the Higgs mechanism and a geometric
contribution $n/R$ from the momentum in the extra dimension.  Their couplings to
the lightest Higgs  and gluons are shown in Fig.  \ref{feynmanrules}. Note that
due to gauge invariance, the coupling of these KK-top quarks to the massless
gluon is identical to that of the SM top quark, while the couplings to the Higgs
boson differ between the SM particles and their KK modes: in our basis choice,
there is a non-mixing contribution which is suppressed by the KK mass, but
yields an SM-like coupling for the zero mode
\begin{equation}
\frac{m^2_{top}}{v \sqrt{\mt^2 + n^2/R^2}} \stackrel{n=0}{\longrightarrow} \frac{\mt}{v} \,,
\end{equation}
as well as a mixing contribution whose coupling converges to the SM yukawa
coupling for {\it large} mode numbers,
\begin{equation}
\frac{m_{top}\frac{n}{R}}{v \sqrt{\mt^2 + n^2/R^2}} \stackrel{n\rightarrow \infty}{\longrightarrow} \frac{\mt}{v} \,.
\end{equation}
These relations will become important in the EFT matching later.  Due to the
mass degeneracy of the two KK top modes, these couplings can in principle be
completely diagonalized if desired, but we choose to keep the basis with mixing
and non-mixing Feynman Rules as given above to keep the hierarchy of suppression
manifest. Let us now turn to the matrix elements for single and double Higgs
production in this model.
\subsection{Single Higgs Production}
The Feynman diagram (up to crossing) contributing to this process in UED is a
straightforward extension of the SM case~\cite{Georgi:1977gs} with an added KK
index,
\begin{center}
 \begin{minipage}[c]{0.5\textwidth}
 \begin{fmffile}{gghed5}
  \begin{fmfgraph*}(72,35)
   \fmfleft{i1,i2}
   \fmfright{o1}
   \fmf{gluon,label=$G_\nu^{b \,, 0}(p_2)$}{i1,v1}
   \fmf{gluon,label=$G_\mu^{a \,, 0}(p_1)$}{i2,v2}
   \fmf{fermion,label.side=right,label=$t_{1,,2}^n$}{v1,v3}
   \fmf{fermion,label=$t_{1,,2}^n$}{v3,v2}
   \fmf{fermion,label.side=right,label.dist=10,label=$t_{1,,2}^n$}{v2,v1}
   \fmf{dashes,label=$h^0$}{o1,v3}
   \fmfv{label.angle=180,label=$\nu$}{v1}
   \fmfv{label.angle=-135,label=$\mu$}{v2}
   \fmfdot{v1,v2,v3}
  \end{fmfgraph*}
 \end{fmffile}
 \end{minipage}
\end{center}
\noindent
Note that the respective contributions of $t_1^n$ and $t_2^n$ to the loop are equal.
\begin{figure}
\centering
 \includegraphics[width=10cm]{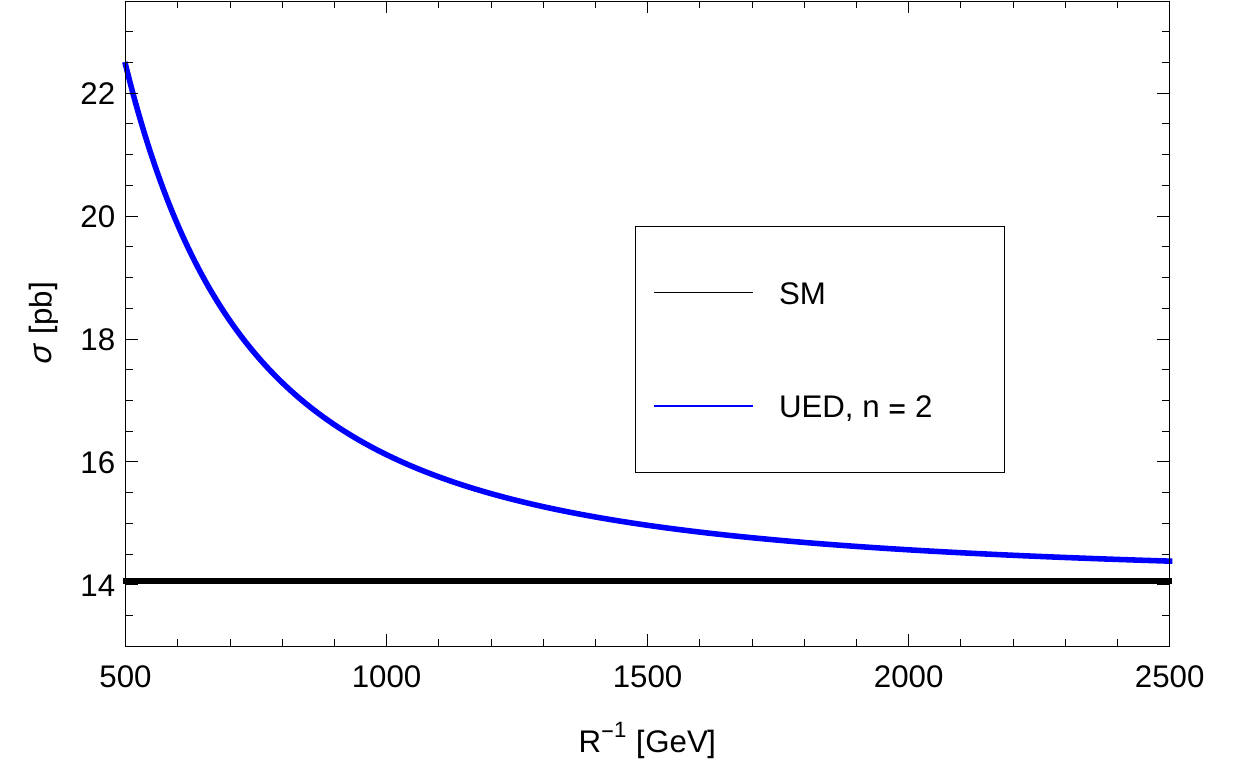}
 \caption{Hadronic LO cross sections for single Higgs production $pp\rightarrow
h$ at $\sqrt{s}=13$ TeV in the SM and in UED including the first two KK-levels
($k \leq 2$). \label{fig1}} 
\end{figure}
We evaluate the matrix elements analytically with FeynCalc \cite{Mertig:1990an},
the scalar integrals are then evaluated numerically with LoopTools
\cite{Hahn:1998yk}. Hadronic cross sections are obtained using the MSTW 2008 NLO
PDFs \cite{Martin:2009iq}.  Figure \ref{fig1} shows the hadronic cross sections
for a proton collider with $\sqrt{s}=13 \tev$ at leading order (LO), contrasting
single Higgs production in the SM with the UED scenario. We consider the first
two KK-top quarks ($n=2$) in the loop and show the dependence of the hadronic
cross section on the inverse radius $R^{-1}$. The cross section is enhanced in
the UED scenario and approaches the SM value from above for large values of
$R^{-1}$ as the KK-top quarks become heavy and decouple from the theory.
\subsection{Higgs Pair Production}
\begin{figure}
\centering
\begin{minipage}[c]{0.3\textwidth}
 \begin{fmffile}{chapgghhmat9}
  \begin{fmfgraph*}(48,30)
   \fmfleft{i1,i2}
   \fmfright{o1,o2}
   \fmf{gluon,label=$G_\nu^b(p_2)$}{i1,v1}
   \fmf{gluon,label=$G_\mu^a(p_1)$}{i2,v2}
   \fmf{fermion,label.side=right,label=$t_{1,,2}^n$}{v2,v1}
   \fmf{fermion,label.side=right,label=$t_{1,,2}^n$}{v3,v2}
   \fmf{fermion,label.side=right,label=$t_{1,,2}^n$}{v4,v3}
   \fmf{fermion,label.side=right,label=$t_{1,,2}^n$}{v1,v4}
   \fmf{dashes,label=$h(p_4)$}{o2,v3}
   \fmf{dashes,label.side=right,label=$h(p_3)$}{o1,v4}
   \fmfdot{v1,v2,v3,v4}
  \end{fmfgraph*}
 \end{fmffile}
  \centering
 $A_{1}$
 \end{minipage}
\hspace{0.5cm}
 \begin{minipage}[c]{0.3\textwidth}
 \begin{fmffile}{chapgghhmat13}
  \begin{fmfgraph*}(48,30)
   \fmfleft{i1,i2}
   \fmfright{o1,o2}
   \fmf{gluon,label.side=left,label=$G_\nu^b(p_2)$}{i1,v1}
   \fmf{gluon,label.side=right,label=$G_\mu^a(p_1)$}{i2,v2}
   \fmf{fermion}{v1,v3}
   \fmf{fermion,label.side=right,label=$t_{1,,2}^n$}{v3,v2}
   \fmf{fermion}{v2,v4}
   \fmf{fermion,label.side=left,label=$t_{1,,2}^n$}{v4,v1}
   \fmf{dashes,label.side=left,label=$h(p_4)$}{o2,v3}
   \fmf{dashes,label.side=right,label=$h(p_3)$}{o1,v4}
   \fmfv{label.angle=-70,label=$t_{1,,2}^n$}{v3}
    \fmfv{label.angle=-80,label=$t_{1,,2}^n$}{v2}
   \fmfdot{v1,v2,v3,v4}
  \end{fmfgraph*}
 \end{fmffile}
 \centering
 $A_{2}$
 \end{minipage}\\ 
 \begin{minipage}[c]{0.3\textwidth}
 \begin{fmffile}{chapgghhmat17}
  \begin{fmfgraph*}(48,30)
   \fmfleft{i1,i2}
   \fmfright{o1,o2}
   \fmf{gluon,label=$G_\nu^b(p_2)$}{i1,v1}
   \fmf{gluon,label=$G_\mu^a(p_1)$}{i2,v2}
   \fmf{fermion,label.side=right,label=$t_{1,,2}^n$}{v2,v1}
   \fmf{fermion,label.side=right,label=$t_{1,,2}^n$}{v3,v2}
   \fmf{fermion,label.side=right,label=$t_{2,,1}^n$}{v4,v3}
   \fmf{fermion,label.side=right,label=$t_{1,,2}^n$}{v1,v4}
   \fmf{dashes,label=$h(p_4)$}{o2,v3}
   \fmf{dashes,label.side=right,label=$h(p_3)$}{o1,v4}
   \fmfdot{v1,v2,v3,v4}
  \end{fmfgraph*}
 \end{fmffile}
 \centering
 $A_{3}$
 \end{minipage}
\hspace{0.5cm} 
 \begin{minipage}[c]{0.3\textwidth}
 \begin{fmffile}{chapgghhmat21}
  \begin{fmfgraph*}(48,30)
   \fmfleft{i1,i2}
   \fmfright{o1,o2}
   \fmf{gluon,label.side=left,label=$G_\nu^b(p_2)$}{i1,v1}
   \fmf{gluon,label.side=right,label=$G_\mu^a(p_1)$}{i2,v2}
   \fmf{fermion}{v1,v3}
   \fmf{fermion,label.side=right,label=$t_{1,,2}^n$}{v3,v2}
   \fmf{fermion}{v2,v4}
   \fmf{fermion,label.side=left,label=$t_{2,,1}^n$}{v4,v1}
   \fmf{dashes,label.side=left,label=$h(p_4)$}{o2,v3}
   \fmf{dashes,label.side=right,label=$h(p_3)$}{o1,v4}
   \fmfv{label.angle=-70,label=$t_{2,,1}^n$}{v3}
    \fmfv{label.angle=-80,label=$t_{1,,2}^n$}{v2}
   \fmfdot{v1,v2,v3,v4}
  \end{fmfgraph*}
 \end{fmffile}
 \centering
 $A_{4}$
 \end{minipage}
\\
  \begin{minipage}[c]{0.3\textwidth}
 \begin{fmffile}{chapgghhmat15}
  \begin{fmfgraph*}(48,30)
   \fmfleft{i1,i2}
   \fmfright{o1,o2}
   \fmf{gluon,label.side=left,label=$G_\nu^b(p_2)$}{i1,v1}
   \fmf{gluon,label=$G_\mu^a(p_1)$}{i2,v2}
   \fmf{fermion,label.side=right,label=$t_{1,,2}^n$}{v1,v3}
   \fmf{fermion,label.side=right,label=$t_{1,,2}^n$}{v3,v2}
   \fmf{fermion,label.side=right,label=$t_{1,,2}^n$}{v2,v1}
   \fmf{dashes,label=$h$}{v4,v3}
   \fmf{dashes,label=$h(p_4)$}{v4,o2}
   \fmf{dashes,label=$h(p_3)$}{v4,o1}
   \fmfdot{v1,v2,v3,v4}
  \end{fmfgraph*}
 \end{fmffile}
 \centering
 $A_{5}$
 \end{minipage}
\caption{The diagrams (up to crossing) contributing to Higgs pair
production. \label{twohiggsdiagrams}} 
\end{figure}
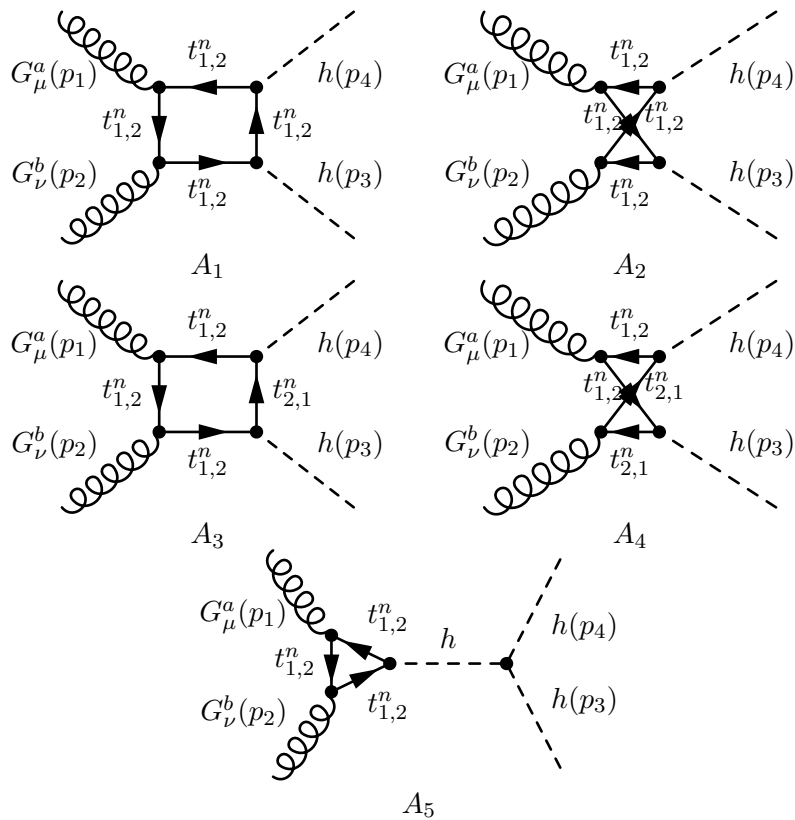

In Higgs pair production one finds a couple of additional diagrams with mixing
vertices compared to the SM case~\cite{Glover:1987nx}.  They are shown in Fig.
\ref{twohiggsdiagrams}.
The first two diagrams ($A_1,A_2$) are topologically distinct non-mixing
diagrams and thus straightforward generalizations of the SM case, while
$A_3,A_{4}$ are a feature of the UED extension as they mix the two top partners
of each massive KK-level. The last diagram ($A_5$) describes production of a
single offshell Higgs splitting into two final state Higgs particles, and again
corresponds to the analogous SM contribution.

\begin{figure}[h]
\centering
 \includegraphics[width=10cm]{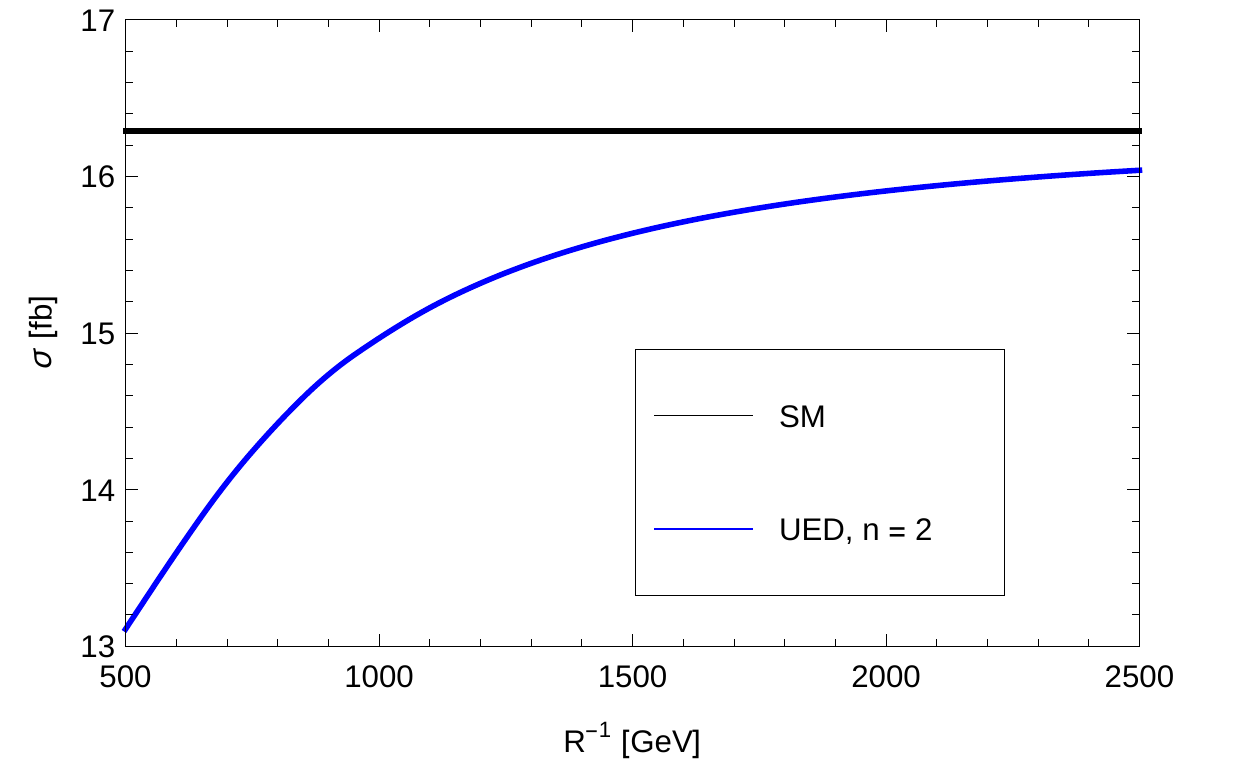}
 \caption{Hadronic cross sections for Higgs pair production $pp\rightarrow h h$
at $\sqrt{s}=13$ TeV in the SM and in UED including the first two KK-levels
($k\leq2$). \label{fig2}} 
\end{figure}

Fig. \ref{fig2} again compares the hadronic cross sections in the SM and in UED
for a proton collider with $\sqrt{s}=13 \tev$. Only the first two excitations
($n=2$) have been taken into account for UED. The renormalization/factorization
scale has been set to twice the Higgs mass $Q=2 \mh$. In contrast to single
Higgs production we observe an $R^{-1}$-dependent reduction of the cross section
compared to the SM.
\section{Effective Theory Description}
\begin{figure}
\begin{center}
\begin{minipage}[c]{0.3\textwidth}
 \begin{fmffile}{effggh1}
  \begin{fmfgraph*}(40,25)
   \fmfleft{i1,i2}
   \fmfright{o1}
   \fmf{gluon,label=$G_\nu^a(p_1)$}{i1,v1}
   \fmf{gluon,label=$G_\mu^b(p_2)$}{v1,i2}
   \fmf{dashes,label=$h(p_3)$}{v1,o1}
   \fmfdot{v1}
  \end{fmfgraph*}
 \end{fmffile}
 \end{minipage}
 \begin{minipage}[c]{0.4\textwidth}
\raisebox{0.1cm}{$\displaystyle -4 i c v \delta_{a b} (p_1 \cdot p_2 g^{\mu \nu} - p_1^\mu p_2^\nu)$}
 \end{minipage}
\\ 
 \vspace{0.5cm} 
 \begin{minipage}[c]{0.3\textwidth}
 \begin{fmffile}{effgghh2}
  \begin{fmfgraph*}(40,25)
   \fmfleft{i1,i2}
   \fmfright{o1,o2}
   \fmf{gluon,label=$G_\nu^a(p_1)$}{i1,v1}
   \fmf{gluon,label=$G_\mu^b(p_2)$}{v1,i2}
   \fmf{dashes,label=$h(p_3)$}{v1,o1}
   \fmf{dashes,label=$h(p_4)$}{o2,v1}
   \fmfdot{v1}
  \end{fmfgraph*}
 \end{fmffile}
 \end{minipage}
 \begin{minipage}[c]{0.4\textwidth}
$\displaystyle -4 i c \delta_{a b} (p_1 \cdot p_2 g^{\mu \nu} - p_1^\mu p_2^\nu)$
 \end{minipage}
\end{center}
\caption{The additional Feynman rules for the $ggh$ and $gghh$ vertices
obtained from the operator $\mathcal O_{GG}$ in
\eqref{mylagrangian}. \label{effectiverules}} 
\end{figure}
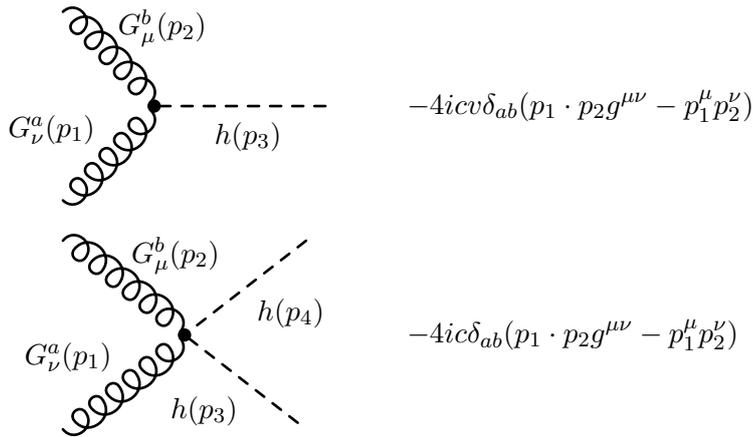

Single on-shell Higgs production in the SM can be described to good precision by
the effective Lagrangian obtained from integrating out the top quark,
\begin{equation}
\sim G_{\mu\nu}^aG^{\mu\nu,a}\,v^2 \ln(1+h/v)=G_{\mu\nu}^a G^{\mu\nu,a} \,\left(v h - \frac{h^2}{2}\right)+O(h^3)\,.
\label{smeffective}
\end{equation}
This effective Lagrangian works well because $\sqrt{\hat s}=m_h<2 m_t$ for
on-shell production.  However, it quickly ceases to be useful for double Higgs
production in which $\sqrt{\hat s}\geq 2 m_h$ very quickly exceeds $2 m_t$,
leading to a breakdown of the heavy top approximation.  New physics in double
Higgs production on the other hand can be sufficiently heavy to allow for a
perturbative effective theory description.  Unlike the (linear) gauge symmetry
breaking top quark contribution \ref{smeffective}, the KK quark modes of our UED
model consistently match to a {\it gauge invariant} dimension-6 operator
\begin{align}
\mathcal O_{GG}= G_{\mu \nu}^a G^{\mu \nu,a} \,
 \left(\phi^\dagger \phi-\frac{v^2}{2}\right)   &=G_{\mu \nu}^a G^{\mu \nu,a} \,\left(v h + \frac{h^2}{2} \right) 
\end{align}
(in unitary gauge) where we have subtracted the vacuum contributions to the
gluon field strength normalization for convenience. Note that in the SM
effective theory in Eq. \ref{smeffective} we find a relative sign between the
single and double Higgs couplings which is absent in the dimension-6 operator.
We normalize the Wilson coefficient of the gauge invariant operator according to 
\begin{equation}\mathcal L=\mathcal
L_{SM}+c\, \mathcal{O}_{GG}\,.\label{mylagrangian}\end{equation}
The resulting Feynman rules are shown in Fig. \ref{effectiverules}.
Higgs pair production in effective theories has recently been studied in
\cite{Goertz:2014qta,Azatov:2015oxa}. Our aim is now to compare this approach to
a concrete new physics scenario.  This issue has been addressed recently for
extended Higgs sectors in \cite{Gorbahn:2015gxa}.  Since both the contributions
to single and double Higgs production originate from the same gauge invariant
operator, we only need to determine the matching constant $c$ once for one of
the 1PI graphs.  We hence carry out the matching procedure for the simpler case
of the $ggh$ coupling and then use the result to consistency check our double
Higgs production calculation. 

Unlike the top quark in the SM effective theory, the KK modes decouple for
$\mkk\rightarrow \infty$ since their Yukawa couplings do not scale like their
masses. It is therefore useful to think of the matching in terms of an expansion
in $\mkk^{-1}$ to the leading nonvanishing order corresponding to an EFT
expansion in $\Lambda^{-1} \sim \mkk^{-1}$, rather than a limit $\mkk\rightarrow
\infty$.  The matching constant naturally depends on the number of
KK-excitations included in the loop.  We obtain
\begin{align}
 c(n) &= \frac{g_{s}^2 \mt^2}{24 \pi^2 v^2} \sum_{k=1}^n \frac{1}{\mkk^2}, \\
 c(\infty) &= \frac{g_s^2}{48 \pi^2 v^2} \left( \pi \frac{\mt}{R^{-1}} \coth \left(\frac{\pi \mt}{R^{-1}} \right) - 1 \right).
\end{align}
While the single and double Higgs interactions obtained from $\mathcal O_{GG}$
only differ by a factor of the vev $v$, the loop integrals in the 1PI-diagrams
contributing to the two processes (namely fermion triangles and boxes
respectively) have respective leading suppression scales $\mkk^{-1}$ and
$\mkk^{-2}$. At first glance, this looks like it might spoil a consistent
matching to a single gauge invariant dimension-6 operator.  This discrepancy is
however resolved by the scaling of the KK mode Yukawa couplings involved: in
triangle diagrams, only non-mixing couplings can appear which themselves are
suppressed with $\mkk^{-1}$, while the box diagrams $A_3,A_4$ contain
unsuppressed mixing couplings. Hence, the leading contributions from triangles
and boxes both scale as $\mkk^{-2}$. This also means that the non-mixing box
diagrams only contribute to dimension-8 operators and do not play a role in the
leading order of EFT.

We use these results to again calculate single and double Higgs production cross
sections, where the SM top quark contributions are now included as loop diagrams
while all contributions from KK modes are absorbed into the effective vertex.

\begin{figure}[h]
\centering
 \includegraphics[width=10cm]{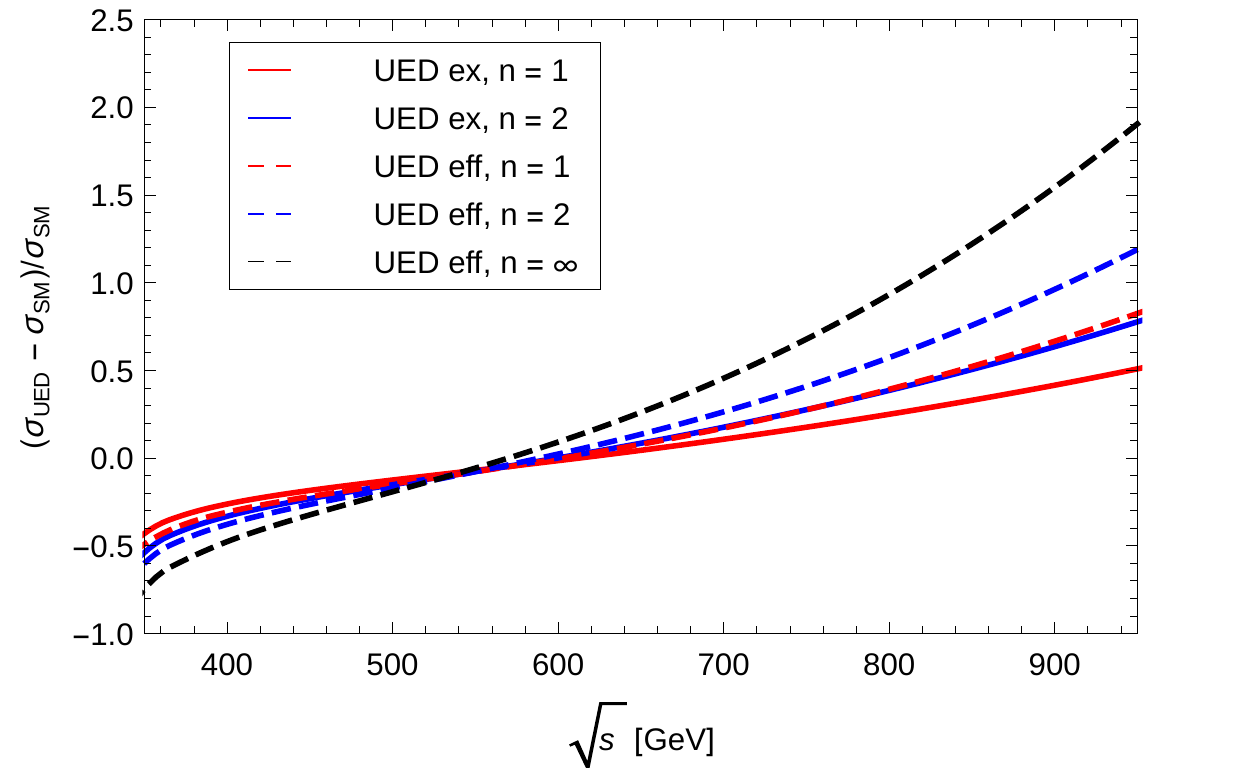}
 \includegraphics[width=10cm]{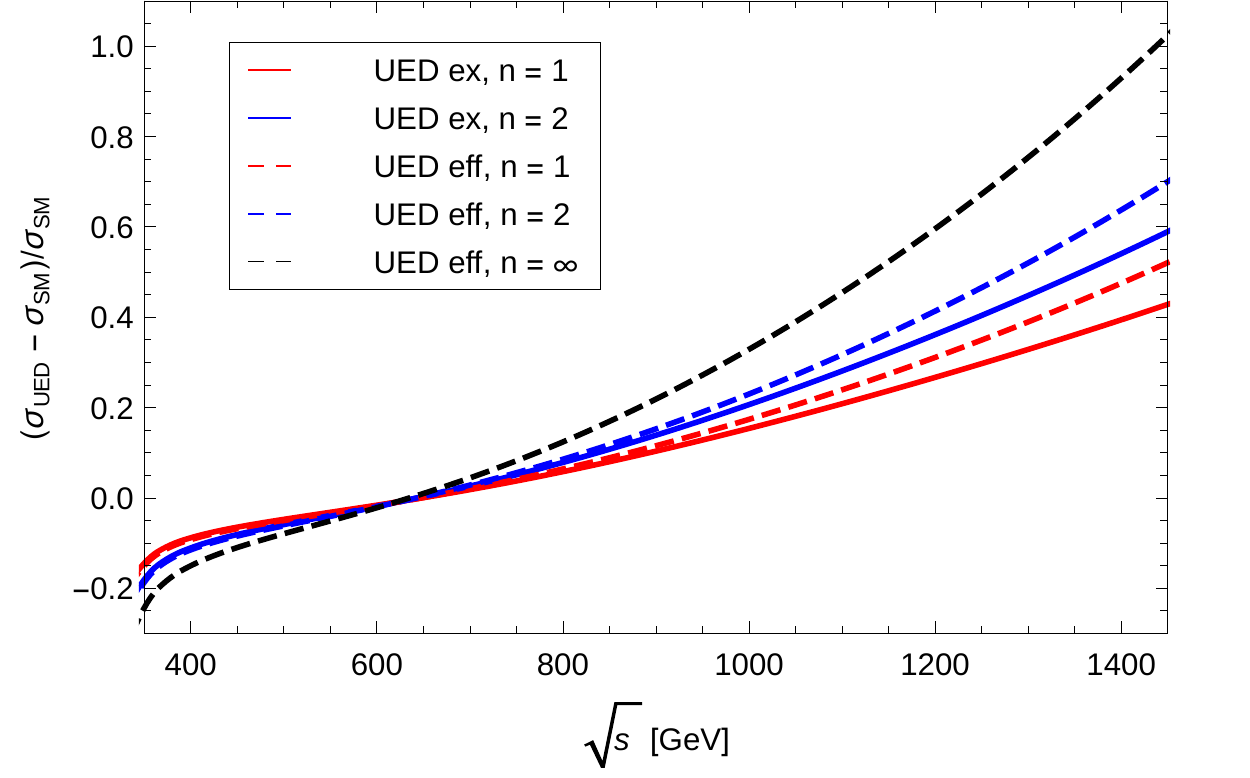}
 \includegraphics[width=10cm]{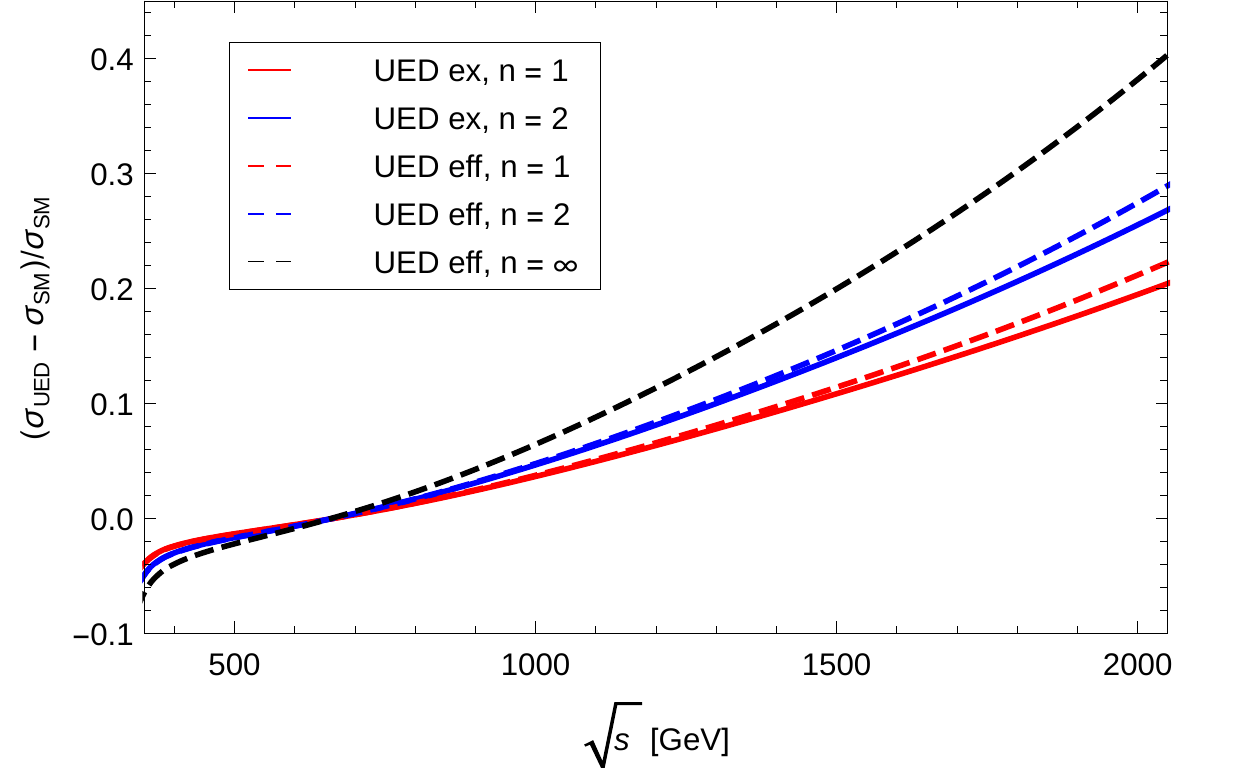}
 \caption{LO partonic double Higgs production cross sections normalized to the
SM. The effects of the KK resonances are taken into account as 1-loop amplitudes
(UED ex, solid) and as effective vertices (UED eff, dashed) respectively for
$R^{-1}=500,1000,2000$ GeV from top to bottom. The result for $n\rightarrow
\infty$ KK-modes is shown in the effective theory case
only.\label{effectivepartonic}}
\end{figure}

\begin{figure}[h]
\centering
 \includegraphics[width=10cm]{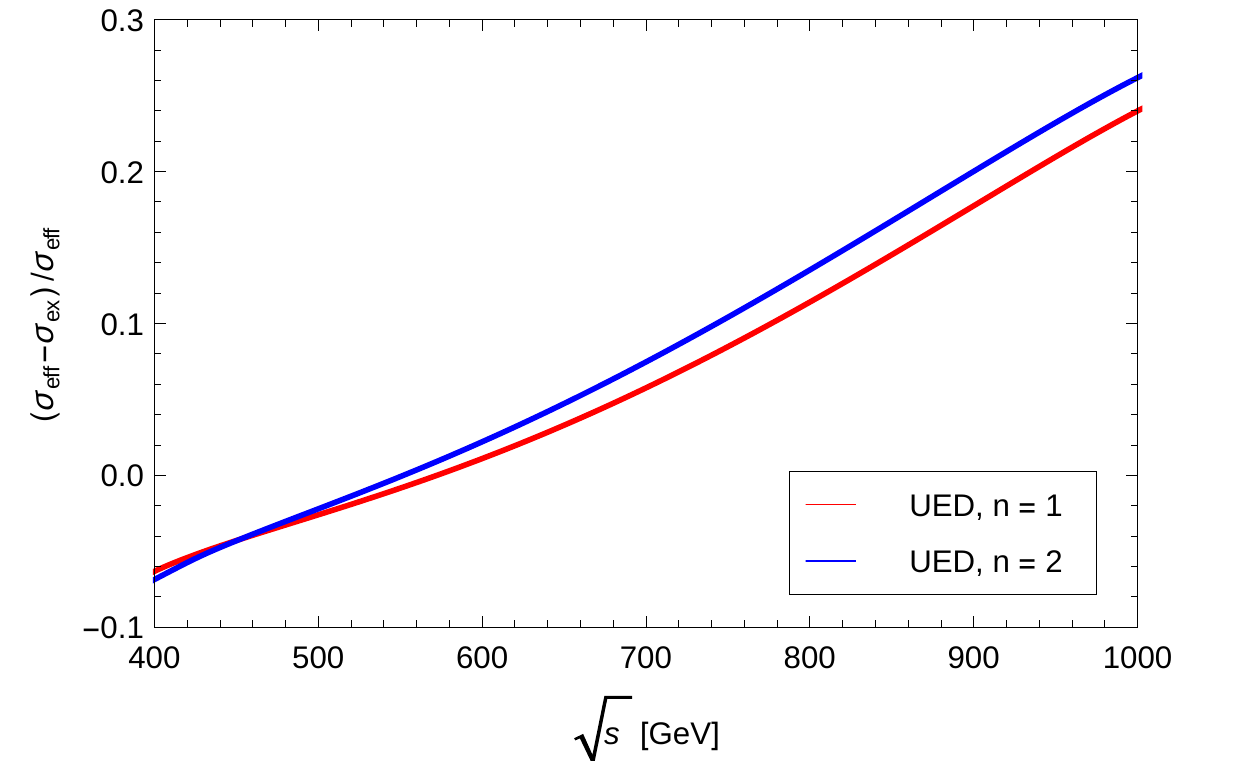}
 \includegraphics[width=10cm]{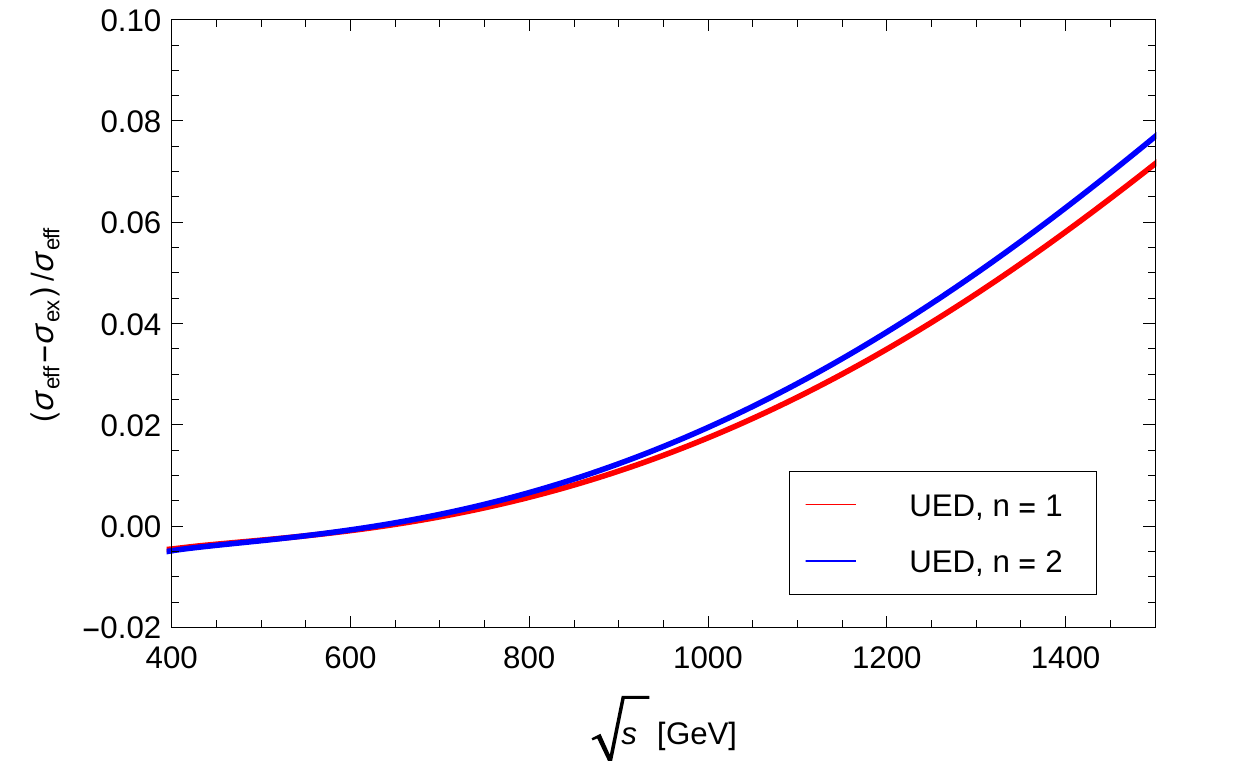}
 \includegraphics[width=10cm]{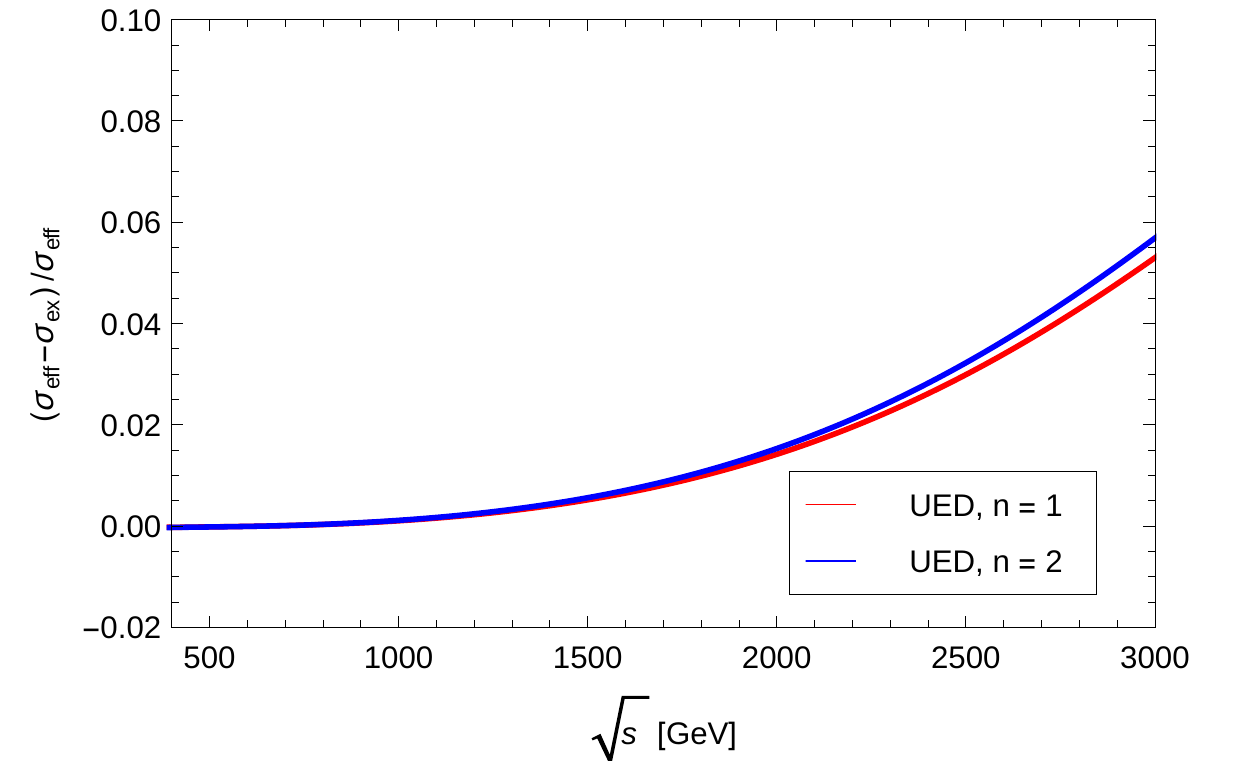}
 \caption{The difference of LO partonic and EFT double Higgs production cross
sections normalized to the EFT with $R^{-1}=500,1000,2000$ GeV from top to
bottom.\label{relativepartonic}}
\end{figure}

Fig. \ref{effectivepartonic} shows the relative deviation $(\sigma_{UED} -
\sigma_{SM})/\sigma_{SM}$ between the SM and UED partonic double Higgs
production cross sections.  The UED KK mode contributions are included as 1-loop
(solid) or EFT amplitudes (dashed).  The simplicity of the EFT approximation
allows to take the limit $n\rightarrow \infty$, which also yields a finite
Wilson coefficient (making this particular process cutoff-independent at LO).
In order to gauge the validity of the EFT approximation, we show a direct
comparison of the 1-loop and effective contributions $\sigma_{ex}$ and
$\sigma_{eff}$ in Fig.~\ref{relativepartonic}.  We notice that the EFT
amplitudes increasingly overestimate the cross section as $\sqrt{\hat s}$
approaches $2\mkk^{(1)}\approx 2R^{-1}$, and underestimate them for $\sqrt{\hat
s}\lesssim 500$ GeV.  This over- and underestimation hence tend to partially
cancel in the total hadronic cross section, but will in principle be visible in
differential distributions.

\begin{figure}[h]
\centering
 \includegraphics[width=10cm]{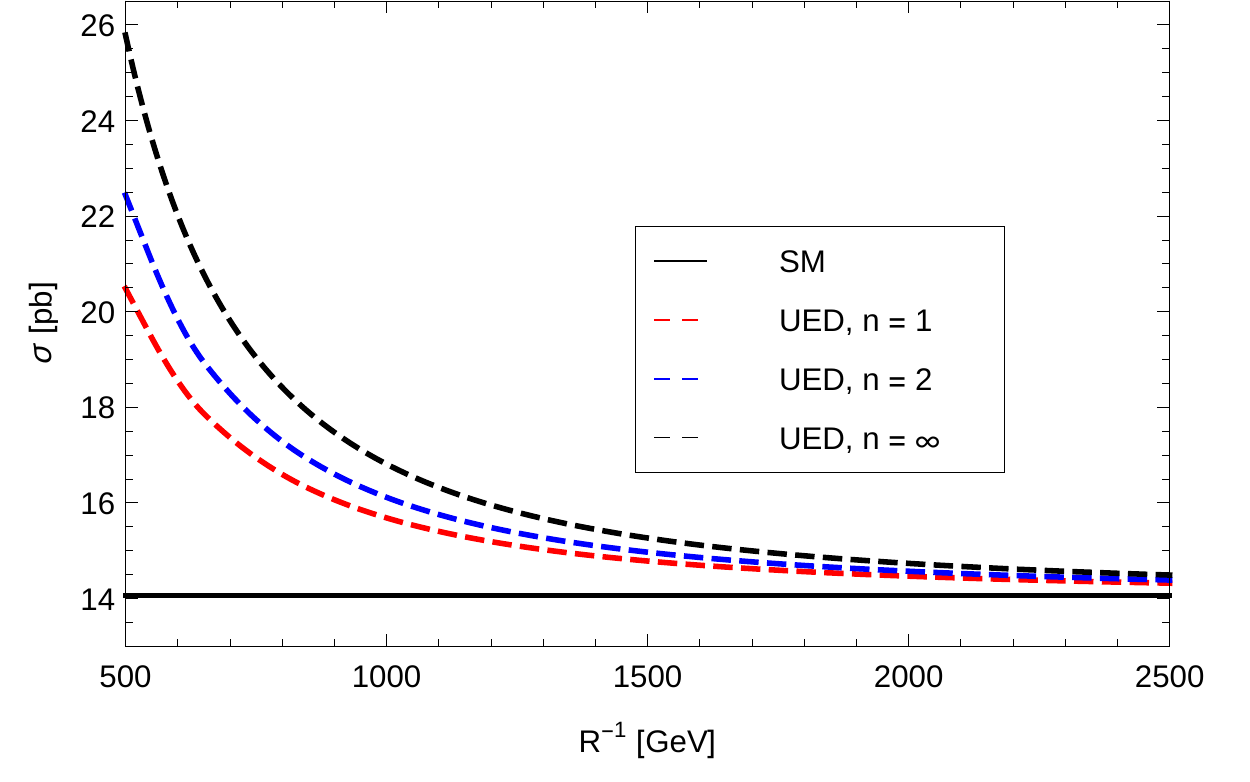}
 \includegraphics[width=10cm]{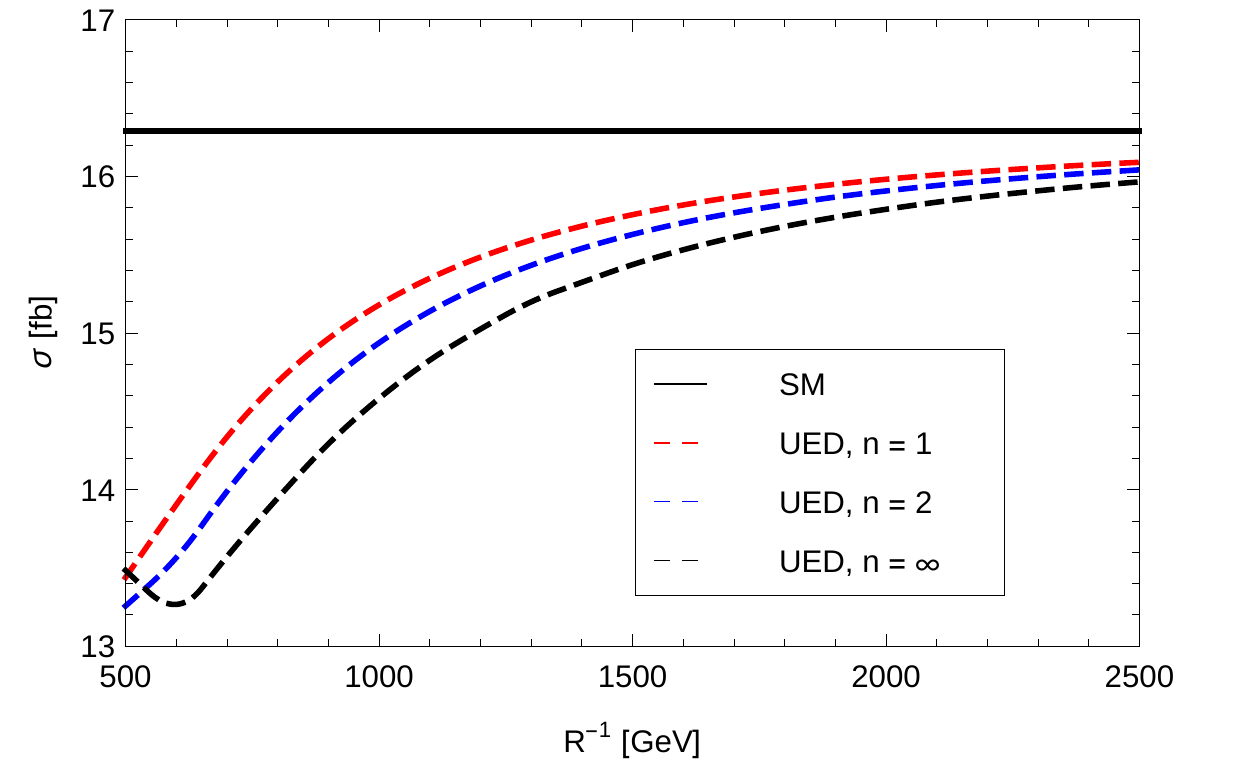}
 \caption{The hadronic cross sections for single and pair production at
$\sqrt{s}=13 \tev$ in the effective description including $n=1,2,\infty$
KK-excitations. \label{effectivehadronic}} 
\end{figure}

Fig. \ref{effectivehadronic} shows the hadronic single and pair production cross
sections for $\sqrt{s} = 13 \tev$ as a function of $R^{-1}$.  Here we compare
the cross sections from the effective description to the SM cross section. We
observe good agreement between our effective theory and the UED cross sections
shown in Figs.~\ref{fig1} and~\ref{fig2}. 


\FloatBarrier
\section*{Acknowledgements}
We thank M. Kr\"amer and F. Riva for valuable discussions.

\end{document}